\documentclass[aps,preprint,floatfix,showpacs,superscriptaddress]{revtex4}

\usepackage{graphicx}
\usepackage{amsmath}
\usepackage{bm}

\newcommand{\be}{\begin{equation}}
\newcommand{\ee}{\end{equation}}
\newcommand{\bel}[1]{\be\label{#1}}
\newcommand{\re}[1]{Eq.~(\ref{#1})}
\newcommand{\ds}{\displaystyle}
\newcommand{\ov}[1]{\overline{#1}}
\newcommand{\hsp}{\hspace*{1pt}}

\begin{document}

\title
{Possible glueball production
in relativistic heavy--ion collisions}


\author{I.N. Mishustin}

\affiliation{Frankfurt Institute for Advanced Studies,
J.W. Goethe Universit\"{a}t,\\
D--60438 Frankfurt am Main, Germany}

\affiliation{The Kurchatov Institute, Russian Research Center,
123182 Moscow, Russia}

\author{L.M. Satarov}

\affiliation{Frankfurt Institute for Advanced Studies,
J.W. Goethe Universit\"{a}t,\\
D--60438 Frankfurt am Main, Germany}

\affiliation{The Kurchatov Institute, Russian Research Center,
123182 Moscow, Russia}

\author{W. Greiner}

\affiliation{Frankfurt Institute for Advanced Studies,
J.W. Goethe Universit\"{a}t,\\
D--60438 Frankfurt am Main, Germany}

\begin{abstract}
Within a thermal model we estimate possible multiplicities of scalar
glueballs in central Au+Au collisions at AGS, SPS, RHIC and LHC
energies. For the glueball mass in the region 1.5--1.7
GeV, the model predicts on average (per event) \mbox{0.5--1.5}
glueballs at RHIC and 1.5--4 glueballs at LHC energies. Possible
enhancement mechanisms are discussed.
\end{abstract}
\maketitle

Glueballs are probably the most unusual particles predicted by the QCD
but not found experimentally yet. It is believed that they can be
produced in ''gluon--rich'' processes, like the $J/\psi$ radiative
decay or $p\ov{p}$ collisions~\cite{Kle04}. The lowest glueball states
with quantum numbers $J^{PC}=0^{++}, 2^{++}, 0^{-+}$ predicted by the
lattice calculations~\cite{Mor99,Mic03} lie in the range 1.4\hsp --\hsp
2.4 GeV~\cite{Mic03}:
\bel{mest}
m\hsp (0^{++})=1.6\pm 0.2\,{\rm GeV},~m\hsp (2^{++})=2.2\pm 0.1\,{\rm GeV},
~m\hsp (0^{-+})=2.4\pm 0.05\,{\rm GeV}\,.
\ee
The uncertainty in masses is mainly introduced by the mixing with the
$q\ov{q}$ states. For example, it is expected that the
$f_0\hsp (1500)$ meson has a very large glueball component (see recent
discussion in Ref.~\cite{Gia05}). We think that relativistic
heavy--ion collisions might be a right tool to produce and study these
exotic particles. If the quark--gluon plasma is indeed produced in such
collisions, a significant fraction, more than 30\%, of its entropy
should be represented by thermal gluons. If these gluons survive until
the hadronization stage, they will inevitably form glueballs as
two--gluon bound states.

It is well established now~\cite{Bec04,And04,And05} that ratios of
hadron multiplicities, observed in central heavy--ion collisions in a
broad range of bombarding energies, can be well reproduced within a
simple thermal model. This model assumes that all hadrons are formed
in a common equilibrated system characterized by the temperature $T$,
the chemical potential $\mu_B$
and the volume $V$. The number density~$N/V$ of bosons in a
baryon--free system is given by the formula ($\hbar=c=1$)
\bel{mult}
\frac{N}{V}=\frac{(2J+1)}{(2\pi)^3}\hspace*{2pt}\gamma_s^{n_s}
\hspace*{-3pt}\int\hspace*{-2pt}{\rm d}^3p
\hsp\left[\exp{\left(\frac{\sqrt{m^2+p^2}}{T}\right)}-1
\right]^{-1}\,,
\ee
where $m$ and $J$ are, respectively, the boson mass and spin.
Following Refs.~\cite{Bec04,Let99} we take into account possible
deviations from chemical equilibrium for hadrons, containing nonzero
number ($n_s$) of strange quarks and antiquarks, by introducing a
strangeness suppression factor $\gamma_s$.

In our previous paper~\cite{Mis05} we have used this model to predict
possible yields of exotic baryonia. In this letter we apply the same
model with the same parameters to estimate multiplicities of
glueballs, assuming that they are in thermal and chemical equilibrium
with other hadrons. In order to eliminate unknown volume we consider
the ratio of the glueball multiplicity to the multiplicity of
$\phi\hsp (1020)$ mesons ($n_s=2$). Assuming further that the glueball is a
flavor--neutral particle we get for this ratio:
\bel{multr}
\frac{N_G}{N_\phi}\simeq\frac{2J+1}{3}\hsp\hsp\gamma_s^{-2}
\left(\frac{m_G}{m_\phi}\right)^{3/2}\hspace*{-2pt}
e^{~\displaystyle (m_\phi-m_G)/T}\,,
\ee
where $m_G\,(m_\phi)$ is the glueball ($\phi$--meson) mass. The r.h.s.
of~Eq.~(\ref{multr}) is obtained in the Boltzmann approximation and in
the lowest order in $T/m$\,. Below we consider only the scalar glueballs
($J=0$).

\begin{table}[ht]
\begin{center}
\caption{\label{tab1}
Parameters of thermal model for central Au+Au and Pb+Pb collisions at
different c.m. energies and the observed multiplicities of
$\phi$--mesons.}
\vspace*{2mm}
\begin{tabular}{c|c|l|l|l}
\hline
reaction & $\sqrt{s_{\scriptscriptstyle NN}}$\,(GeV) &~$T$\,(MeV)~&~$\gamma_s$~&~$N_\phi$~\\
\hline
Au+Au&~\small 4.87&~\small 119.1~\cite{Bec04}&~\small 0.763~\cite{Bec04}&~\small 1.5\hsp $\pm$\hsp  0.3~\cite{Bac04,Ahl99}\\
Pb+Pb&~\small 8.87&~\small 145.5~\cite{Bec04}&~\small 0.807~\cite{Bec04}&~\small 2.57\hsp $\pm$\hsp 0.1~\cite{Afa00}\\
Pb+Pb&~\small 12.4&~\small 151.9~\cite{Bec04}&~\small 0.766~\cite{Bec04}&~\small 4.37\hsp $\pm$\hsp 0.14~\cite{Afa00}\\
Pb+Pb&~\small 17.3&~\small 154.8~\cite{Bec04}&~\small 0.938~\cite{Bec04}&~\small 7.6\hsp $\pm$\hsp 1.1~\cite{Afa00}\\
Au+Au&~\small 130&~\small   176~\cite{And04}&~\small   1.0~\cite{And04}&~\small 34\hsp $\pm$\hsp 5~\cite{Adl02}\\
Au+Au&~\small 200&~\small   177~\cite{And04}&~\small   1.0~\cite{And04}&~\small 38\hsp $\pm$\hsp 3~\cite{Adl05,Bea05}\\
Au+Au&~\small $5.5\cdot 10^3$&~\small   177&~\small   1.0&\\
\hline
\end{tabular}
\end{center}
\end{table}
We use the parameters of thermal model obtained in Refs.~\cite{Bec04}
(fit B) and~\cite{And04} by fitting the hadron ratios observed in
central Au+Au and Pb+Pb collisions at various energies. These
parameters as well as experimental values of the $\phi$--meson
multiplicities in these reactions are given in Table~\ref{tab1}. In the
case of Au+Au collisions at the AGS and RHIC energies the total $\phi$
multiplicities are not yet available. For these reactions we use the
$\phi/\pi$ ratios observed at midrapidity~\cite{Bac04,Adl02,Adl05} and
estimate $N_\phi$ multiplying these ratios by charged pion
multiplicities~\cite{Ahl99,Bea05} extrapolated to the whole rapidity
space. At the LHC energy $\sqrt{s_{NN}}=5.5$\,TeV we take the same $T$
and $\gamma_s$ as for central Au+Au collisions at
$\sqrt{s_{NN}}=200$\,GeV and use the value $N_\phi=100$, obtained by
extrapolating the RHIC data. In this extrapolation we assume that the
ratio of $\phi$ multiplicities at the LHC and RHIC energies equals the
corresponding ratio of charged particle multiplicities,~$N_{ch}$. The
phenomenological formula for the energy dependence of $N_{ch}$ is taken
from Ref.~\cite{Bac05}.

\begin{figure*}[htb!]
\vspace*{-8cm}
\hspace*{1cm}\includegraphics[width=0.8\textwidth]{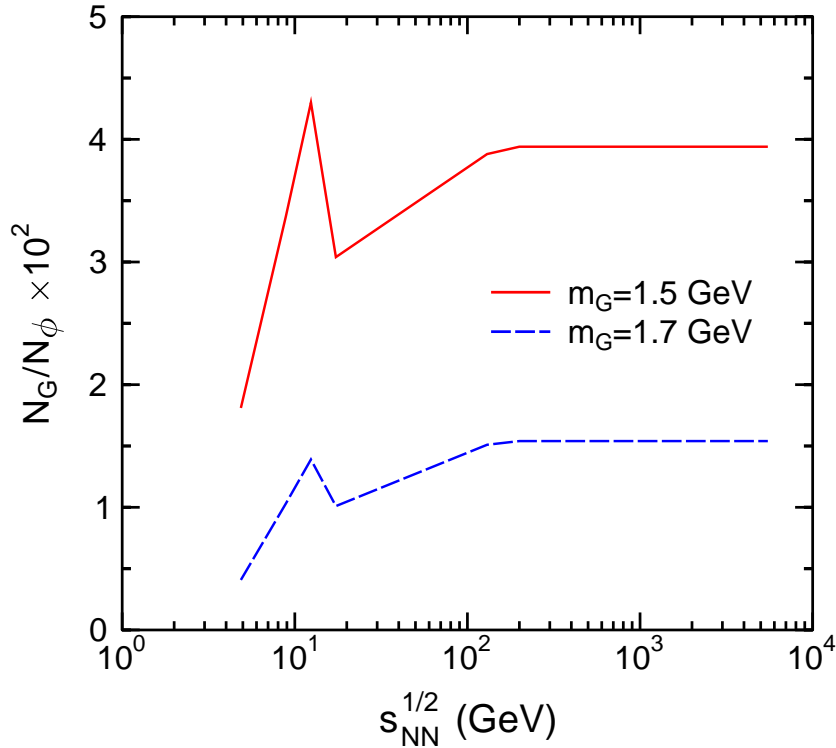}
\caption{Ratio of mean multiplicities of glueballs and
$\phi$--mesons in central Au+Au collisions at different c.m. bombarding
energies calculated within the thermal model. Solid and dashed lines
show results for glueballs with masses $m_G=1.5$ and 1.7 GeV, respectively.}
\label{fig1}
\end{figure*}
By using \re{multr} with the parameters from Table~\ref{tab1} we
calculate $N_G/N_\phi$ for the case of most central Au+Au collisions at
different bombarding energies. The results for two values of the
glueball mass, $m_G=1.5$ and 1.7 GeV, are shown in Fig.~\ref{fig1}.
A nontrivial behavior of the $N_G/N_\phi$--ratio is explained by the
nonmonotonic energy dependence of~$\gamma_s$. One can see that for the
most central Au+Au collisions at $\sqrt{s_{NN}}\gtrsim 200$\,GeV this
ratio is predicted in the range of 1.5\hsp --\hsp 4\%.

\begin{figure*}[htb!]
\vspace*{-8cm}
\hspace*{1cm}\includegraphics[width=0.8\textwidth]{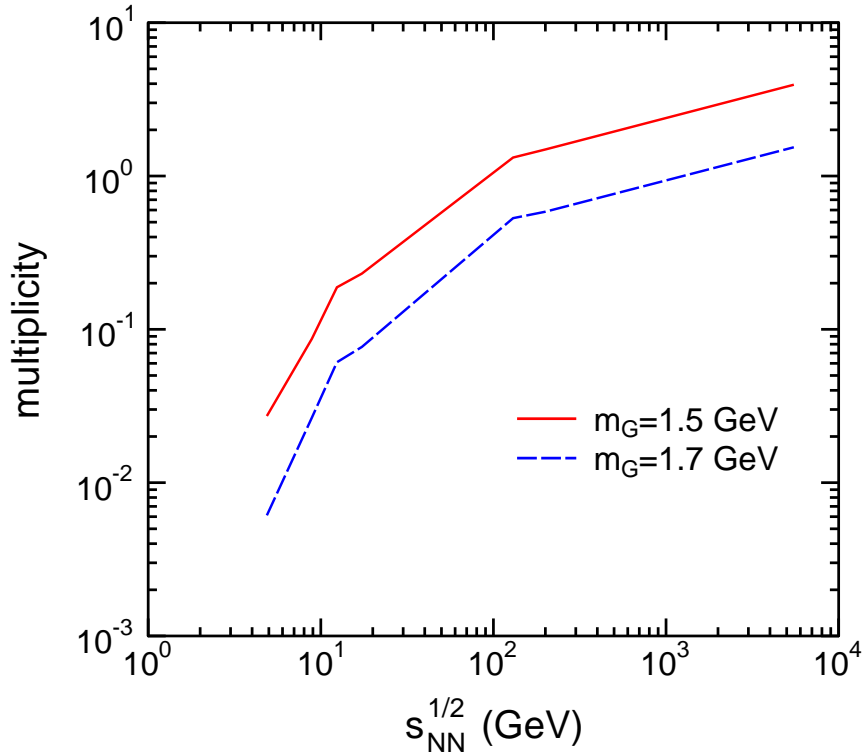}
\caption{Mean multiplicities of glueballs
in central Au+Au collisions
for different values of mass~$m_G$\,.
}
\label{fig2}
\end{figure*}
Figure~\ref{fig2} shows the excitation functions of the absolute
glueball multiplicity $N_G$ in the central Au+Au collisions.
At higher RHIC energy the predicted
glueball yields are in the range of 0.6\hsp --\hsp 1.5 (per event) for
$m_G=$\hsp 1.5\hsp --\hsp 1.7 GeV. At the future LHC facility we
predict, respectively, 1.5--4 glueballs per central Au+Au collision.

It is instructive to compare these predictions with glueball
multiplicities obtained for superposition of independent nucleon--nucleon
collisions. In order to estimate glueball yields in a single
NN--collision we use the thermal model of $pp$ and $p\ov{p}$ reactions,
suggested in Ref.~\cite{Be04a}. Within this model, the hadron
multiplicities observed in $p\ov{p}$ interactions at
$\sqrt{s}=200$\,GeV can be reproduced with the effective
temperature $T\simeq 175\pm 11$\,MeV and the particle emitting volume
$V\simeq 35\pm 14$\,fm$^3$\,. Using~\re{mult}, we get the following
estimate for average multiplicity of glueballs in a single
$p\ov{p}$--collision at \mbox{$\sqrt{s}=200$\,GeV}:
\bel{est}
N_G\,({p\ov{p}})\simeq (3.4-9.0)\cdot 10^{-3}\,.
\ee
The two values in the r.h.s. correspond to $m_G=1.5$\,and 1.7~GeV. We
further assume that probabilities of the glueball production in the
$p\ov{p}$-- and the NN--collisions are the same at high energies.
Multiplying \re{est} by $N_{\rm part}/2\simeq 178$~\cite{Bea04}, the
average number of NN--pairs participating in most central Au+Au
collisions, we obtain 0.6-1.6 glueballs per central Au+Au collision at
$\sqrt{s_{NN}}=200$\,GeV. From Fig.~\ref{fig2} one can see that these
values are rather close to the glueball multiplicities, predicted by
the thermal model for the highest RHIC energy.

We think, that the estimates presented above should be regarded as a lower
bound for the glueball yield in high--energy heavy--ion collisions.
As follows from the QCD--based calculations~\cite{Kar97,Lev98}
and also from some effective models~\cite{Mon98,Car00},
the gluon--like excitations acquire a temperature--dependent mass,
$m_g\simeq 0.8$\,GeV just before the hadronization transition.
Such quasiparticles may form glueball--like bound states
already in the plasma phase~\cite{Shu04}.
Applying the same formula (\ref{mult}) for gluonic
quasiparticles and replacing $2J+1$ by the effective statistical weight
of massive gluons, $\nu_g\sim 10$, one gets the ratio of the
scalar glueball multiplicity to the gluon multiplicity at $T=T_c$:
\bel{ggr}
\frac{N_G}{N_g}\simeq
\nu_g^{-1}\left(\frac{m_G}{m_g}\right)^{3/2}
\hspace*{-3pt}e^{~\displaystyle (m_g-m_G)/T_c}\sim 0.3\%\,,
\ee
where the numerical value is obtained assuming $T_c=170$\,MeV and
$m_G=2m_g$\,.

However, in a rapidly expanding system the abundances of different
species will be determined not only by the temperature, but also by
the corresponding reaction rates. In this case the $N_G/N_g$--ratio may
significantly overshoot the ''equilibrium''
value~(\ref{ggr})\hsp\footnote
{
Analogous mechanism leading to the overpopulation of
$q\ov{q}$--pairs was proposed in Ref.~\cite{Let99}.
}.
Indeed, during the hadronization, the massive gluons may decay
into \mbox{$q\ov{q}$--pairs} or recombine into the $gg$--bound states
forming, respectively, mesons or glueballs.
Using simple kinetic equations for
glueball and gluon abundances with the recombination term, proportional
to the $gg\to G$ cross section, $\sigma_{gg}$, and the $g\to q\ov{q}$
decay term, characterized by the width $\Gamma_g$\,, we obtain the
estimate
\bel{gneq}
\frac{N_G}{N_g}\sim\frac{\ds n_g\sigma_{gg}v_{\rm rel}}{\ds\Gamma_g}
\sim 5\%\,.
\ee
Here we have used the  typical values \mbox{$\sigma_{gg}\sim 10$\,mb},
\mbox{$\Gamma_g\sim 100$\,MeV},
\mbox{$v_{\rm rel}\sim \sqrt{T_c/m_g}\simeq 0.5$} and the equilibrium
gluon density $n_g=n_g\hsp (T_c)$ \mbox{$\simeq 0.05$\,fm$^{-3}$}.
The ratio (\ref{gneq}) exceeds the thermal estimate (\ref{ggr})
by more than one order of magnitude.
Therefore, much more favorable conditions for the glueball production
may be realized in the explosive hadronization scenario~\cite{Mis99}.

For the experimental identification of glueballs one can use the decay
mode \mbox{$G\rightarrow K\overline{K}$} with predicted partial width
in the range 10\hsp --\hsp 40 MeV. For typical conditions, when
glueballs are slow in the c.m. frame, one should look for the
back--to--back correlated kaons with individual momenta of about
0.6\hsp --\hsp 0.7 GeV/c. The \mbox{$G\to\gamma\gamma$} channel with
the partial width of about 1-10 keV~\cite{Gia05} can also be used for the
glueball identification.

In conclusion, we have used a simple thermal model to predict possible
yields of scalar glueballs in central Au+Au collisions at different
energies, from AGS to LHC. Their maximal multiplicities are predicted in the
range of 1.5-4\% of the $\phi$--meson multiplicity. Even larger yields are
expected in the case of explosive hadronization of the quark--gluon
plasma. We believe that such yields are in the reach of experimental
observations.

\begin{acknowledgments}
This work has been partly supported by the GSI (Germany),
the DFG Grant \mbox{436 RUS 113/711/0-2},
the RFBR Grant 05--02--04013,
and the MIS Grant \mbox{NSH--8756.2006.2}.
\end{acknowledgments}

\end{document}